# MACH, THE UNIVERSE, AND FOUNDATIONS OF MECHANICS

B. Mashhoon[1] and P.S. Wesson[2,3]

[1]Department of Physics and Astronomy, University of Missouri, Columbia, Missouri 65211, U.S.A.

[2]Department of Physics and Astronomy, University of Waterloo, Waterloo, Ontario N2L 3G1, Canada

[3]Herzberg Institute of Astrophysics, National Research Council, Victoria, B.C. V9E 2E7, Canada

Abstract

Barbour's response to our recent paper on "Mach's principle and higher-dimensional dynamics" describes an approach to Mach's principle in which the universe as a whole is involved in the definition of inertial frames of reference. Moreover, Barbour's theoretical procedure is in agreement with general relativity for a finite universe that is spatially closed. However, we prefer an operational approach that relies ultimately on observational data.

Corresponding author: B. Mashhoon (e-mail address: MashhoonB@missouri.edu).

In his response [1] to our recent paper [2] regarding Mach's principle, Barbour has briefly explained his definition of Mach's principle and how general relativity is in his sense Machian. As we have already acknowledged [2], other approaches to Mach's principle do exist that are different from ours.

In Barbour's approach, the universe as a whole plays a crucial role in the foundations of mechanics. To clarify this point, let us return to Mach's discussion of the operational definition of inertial frames of reference [3] that we quoted in section IV of our paper [2] and recall Mach's argument that space and time as physical concepts are distinct from their operational definitions by means of masses. This important point hinges on the separation between "local" astronomical masses that are used to define time and position operationally and *the rest of the matter in the universe*. The *only* way to avoid Mach's conclusion is to assume that *all* of the masses in the universe are employed in the definition of inertial frames of reference. This would be impossible in practice, of course, but would turn Mach's dream of the knowledge of "*immediate* connections" of all the masses in the universe into a conceptual challenge.

Barbour has described a possible theoretical procedure in which the whole universe could be employed to define inertial frames of reference [1]. After placing certain constraints on the universe, namely, that the universe is finite and spatially closed, Barbour concludes that general relativity would be compatible with this approach to Mach's principle. This end result of Barbour's extensive theoretical work harks back to earlier suggestions by Einstein, Wheeler, and others [4, 5] that Mach's principle implied a finite spatially bounded universe as a plausible way of avoiding unascertainable boundary conditions in cosmology.

To conclude, in Barbour's approach to Mach's principle the whole inaccessible universe plays a crucial role in the formulation of foundations of experimental science. Moreover, compatibility with general relativity is achieved for a spatially closed universe. In this connection, we prefer to rely on the judgment of observational cosmology.